\documentclass[english]{article}
\usepackage[T1]{fontenc}
\usepackage[latin9]{inputenc}
\usepackage{xcolor}
\usepackage{pdfcolmk}
\usepackage{calc}
\usepackage{amstext}
\usepackage{graphicx}
\usepackage[all]{xy}
\PassOptionsToPackage{normalem}{ulem}
\usepackage{ulem}

\makeatletter

\providecolor{lyxadded}{rgb}{0,0,1}
\providecolor{lyxdeleted}{rgb}{1,0,0}

\DeclareRobustCommand{\lyxsout}[1]{\ifx\\#1\else\sout{#1}\fi}

\newcommand{\lyxaddress}[1]{
	\par {\raggedright #1
	\vspace{1.4em}
	\noindent\par}
}

\@ifundefined{date}{}{\date{}}
\usepackage{geometry}
\usepackage{cite}

\makeatother

\usepackage{babel}
\begin{document}

\title{Contextuality Analysis of the Double Slit Experiment\\(With a Glimpse
Into Three Slits) }

\author{Ehtibar N. Dzhafarov and Janne V. Kujala}
\maketitle

\lyxaddress{Purdue University, USA, ehtibar@purdue.edu and University of Jyväskylä,
Finland, jvk@iki.fi}
\begin{abstract}
The Contextuality-by-Default theory is illustrated on contextuality
analysis of the idealized double-slit experiment. The experiment is
described by a system of contextually labeled binary random variables
each of which answers the question: has the particle hit the detector,
having passed through a given slit (left or right) in a given state
(open or closed)? This system of random variables is a cyclic system
of rank 4, formally the same as the system describing the EPR/Bell
paradigm with signaling. Unlike the latter, however, the system describing
the double-slit experiment is always noncontextual, i.e., the context-dependence
in it is entirely explainable in terms of direct influences of contexts
(closed-open arrangements of the slits) upon the marginal distributions
of the random variables involved. The analysis presented is entirely
within the framework of abstract classical probability theory (with
contextually labeled random variables). The only physical constraint
used in the analysis is that a particle cannot pass through a closed
slit. The noncontextuality of the double-slit system does not generalize
to systems describing experiments with more than two slits: in an
abstract triple-slit system, almost any set of observable detection
probabilities is compatible with both a contextual scenario and a
noncontextual scenario of the particle passing though various combinations
of open and closed slits (although the issue of physical realizability
of these scenarios remains open).

\end{abstract}

\section{Introduction}

\begin{figure}
\fbox{\begin{minipage}[t]{0.5\columnwidth}%
\includegraphics[scale=0.25]{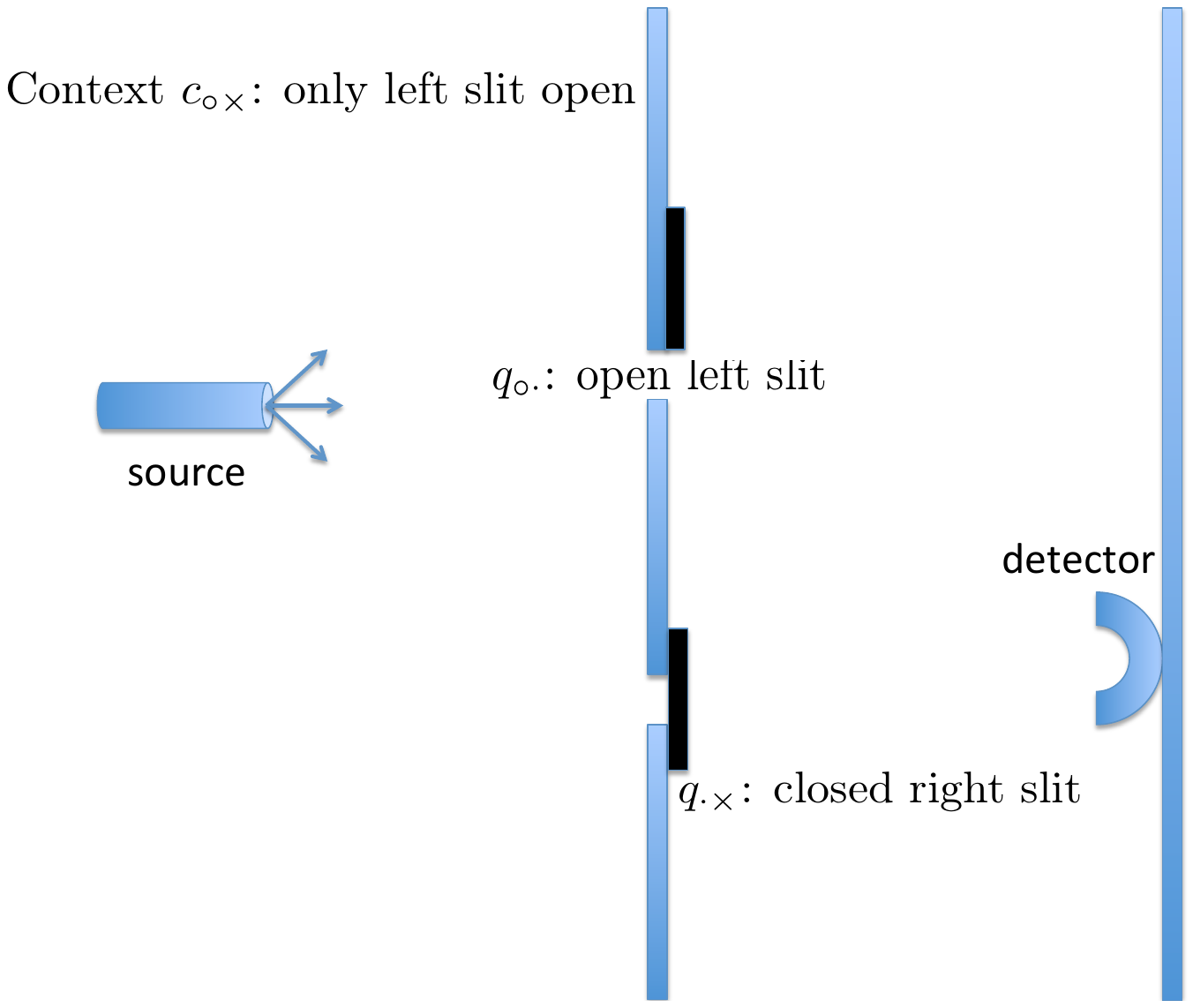}%
\end{minipage}}%
\fbox{\begin{minipage}[t]{0.5\columnwidth}%
\includegraphics[scale=0.25]{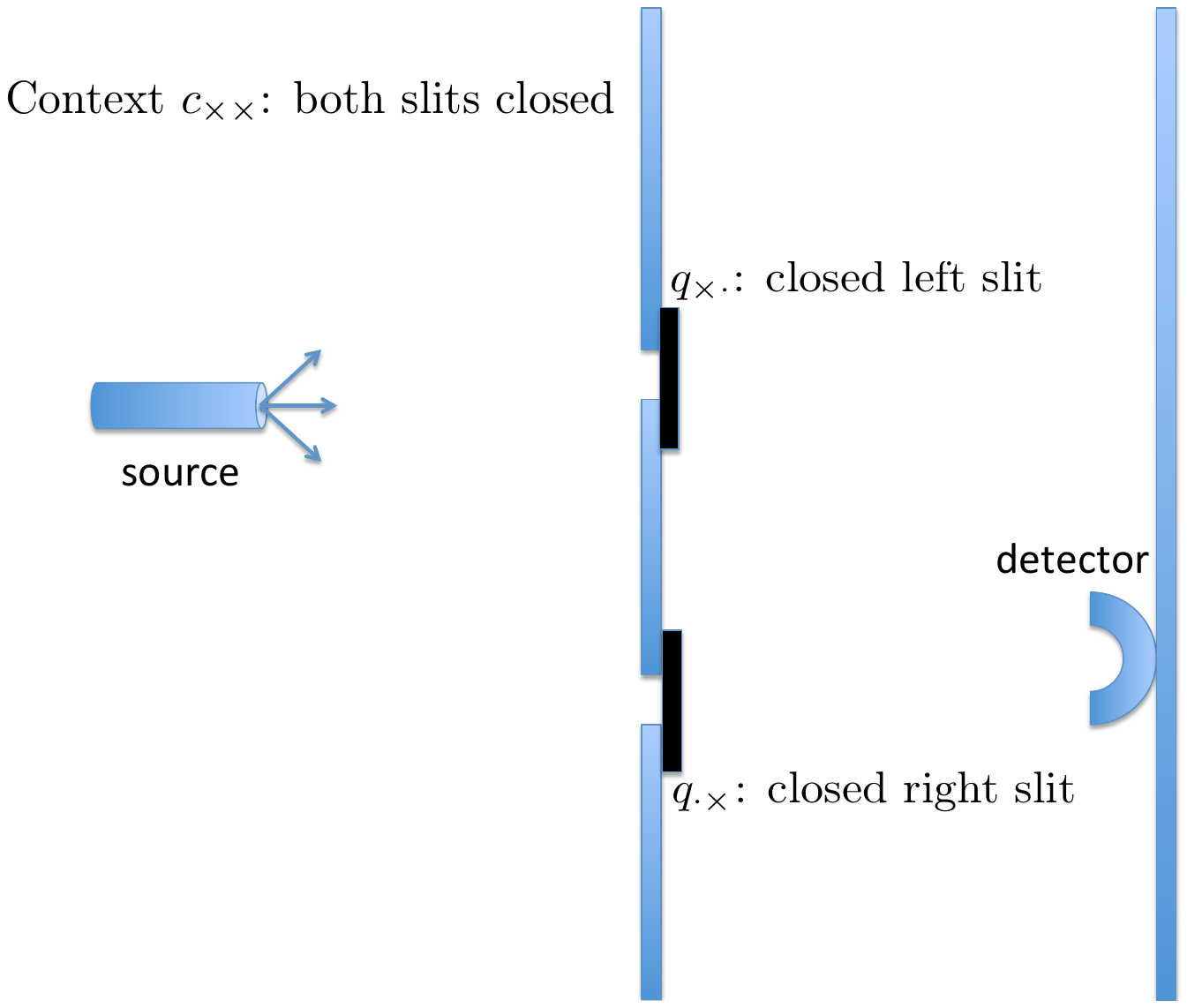}%
\end{minipage}}

\fbox{\begin{minipage}[t]{0.5\columnwidth}%
\includegraphics[scale=0.25]{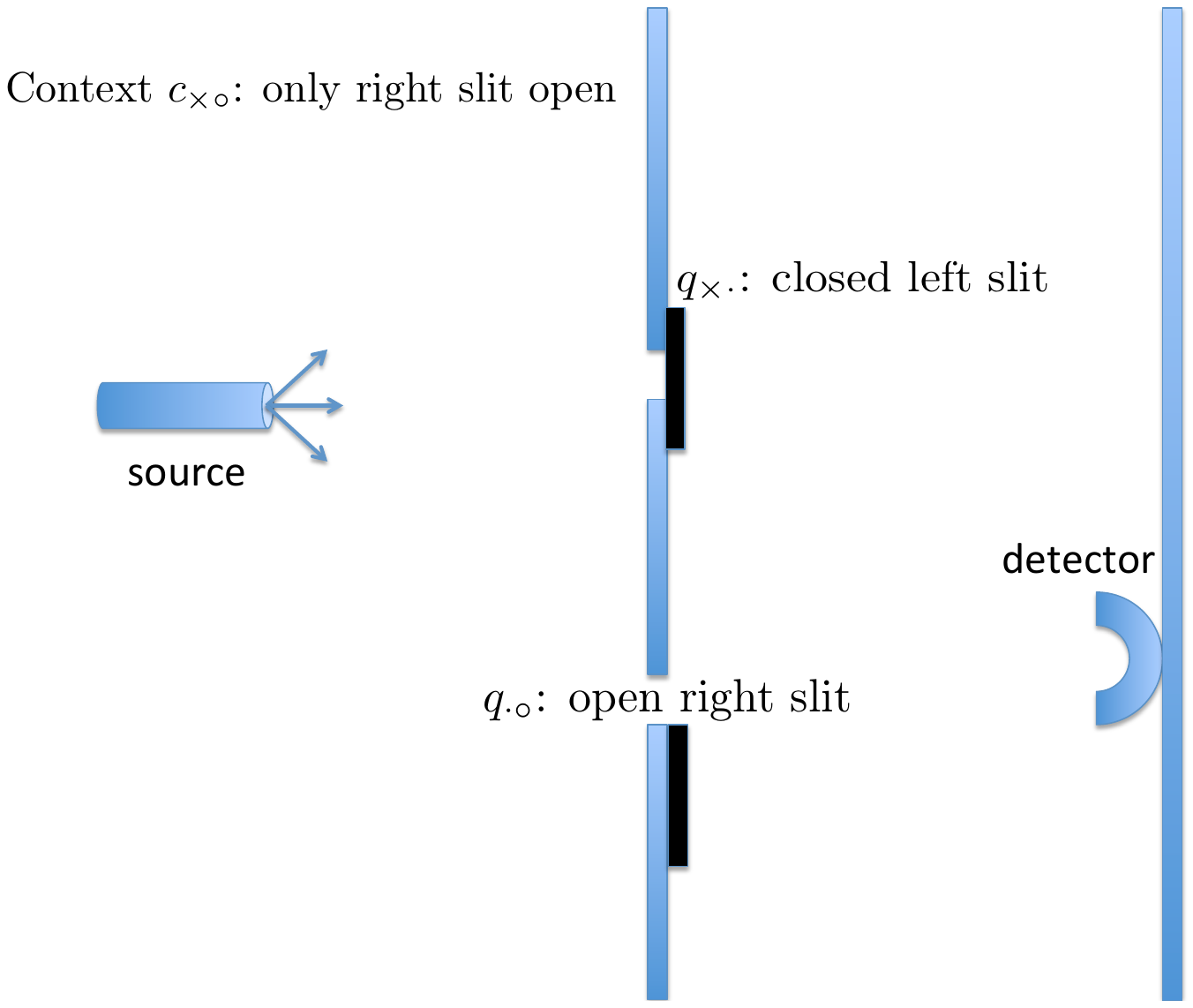}%
\end{minipage}}%
\fbox{\begin{minipage}[t]{0.5\columnwidth}%
\includegraphics[scale=0.25]{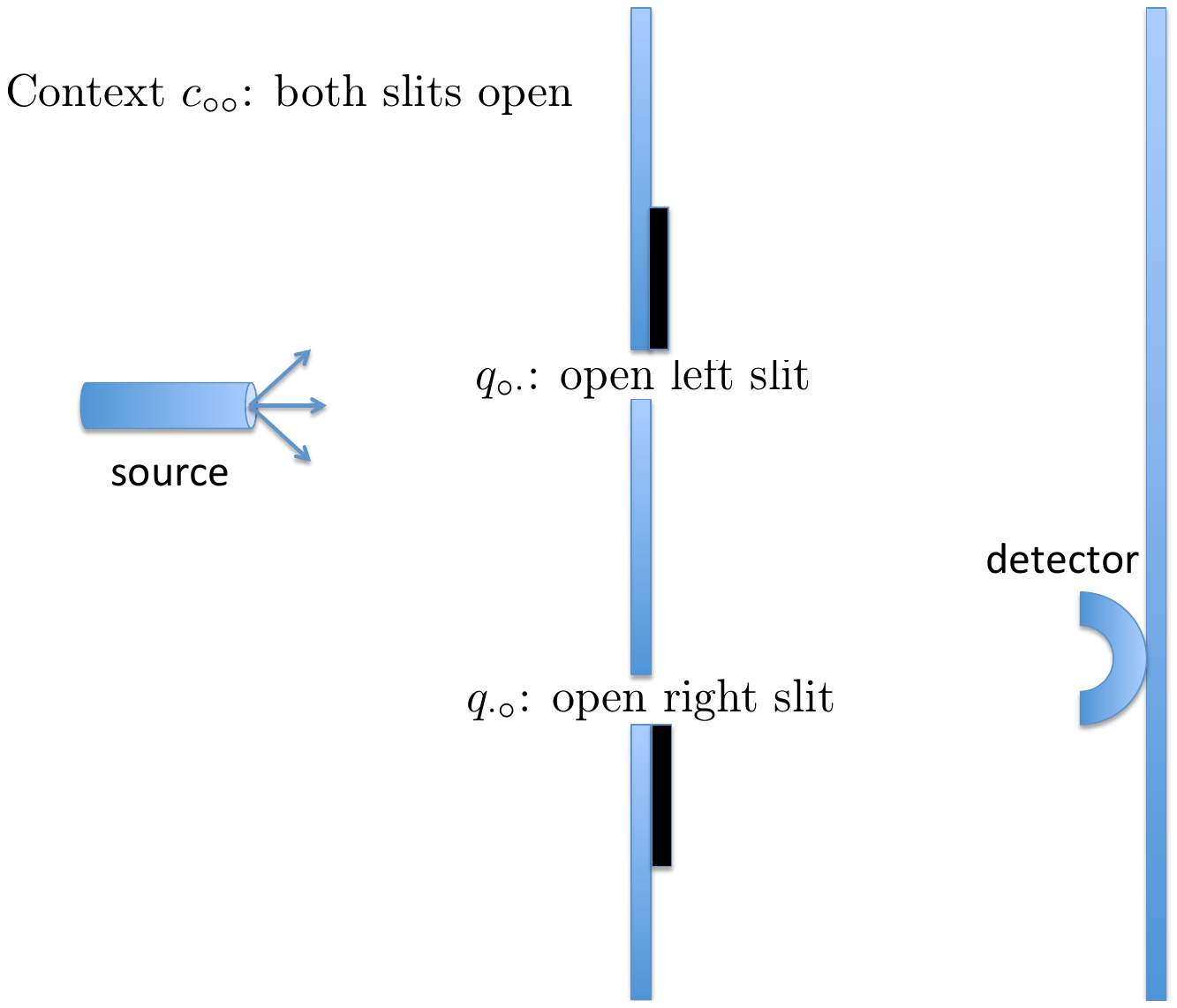}%
\end{minipage}}

\caption{\label{fig: 4 contexts}Idealized double slit experiment. The source
shoots a single particle that may or may not pass through the slits
cut in the intermediate plate, and may or may not be recorded by the
detector placed on the screen behind the plate. Each of the slits
can be open or closed, and each of the four closed-open arrangements
of the two slits forms a context. Each slit (left or right) in each
state (open or closed) forms a content, formally treated as the property
of the physical system measured by a random variable in a given context.}
\end{figure}

This note is an illustration of the workings of the Contextuality-by-Default
(CbD) theory \cite{DzhCerKuj2017,DzhKuj2017Fortsch,KujDzhLar2015,Dzh2017Nothing}
on the classical double-slit experiment. Specifically, we consider
the single-particle version of this experiment, schematically depicted
in Figure$\,$\ref{fig: 4 contexts}, and represent it by a system
of binary random variables $R_{q}^{c}$ answering the question:
\begin{description}
\item [{Q1:}] in context $c,$ has the particle emitted by the source hit
the detector, having passed through slit $q$?
\end{description}
Here, $q$ denotes a particular slit, left or right, and whether it
is open or closed, whereas $c$ denotes the variable part of the experimental
set-up: which of the two slits is open and which is closed. The answer
to the question Q1 is Yes ($R_{q}^{c}=+1)$ if the conjunction of
the following two events occurs: the particle passed through $q$,
and the particle hit the detector. If this has not happened, $R_{q}^{c}=-1$.
For instance, if $c=c_{\circ\times}$ (the left slit is open, the
right one is closed) and $q=q_{\cdot\times}$ (indicating the closed
right slit), then $R_{q_{\cdot\times}}^{c_{\circ\times}}=+1$ means
that the particle passes through the closed right slit and hits the
detector. The probability of this happening is, of course, zero. It
is in fact the only physical assumption used in our analysis: that
it is impossible for a particle to pass through a closed slit. This
can be complemented by the statement (we choose not to consider it
as a separate assumption) that it is meaningful to speak of a particle
passing or not passing through an open slit. We assume nothing else
about possible trajectories, and do not even commit to any specific
meaning of the term ``trajectory'' (the graphical illustrations
in Figs. \ref{fig: cox}-\ref{fig: coo} being merely visual aids).
We allow the particle to pass through more than one slit at a time,
any number of times and in any succession or simultaneously, before
hitting the detector or missing it.

We do not set detectors at the slits (which would, as is well known,
dramatically change and constrain possible outcomes of the experiment).
The only recordable event in our analysis is whether at the end of
the experiment the detector placed in the receiving plane has been
hit or missed. For any probability of this happening we have therefore
to consider all possible scenarios of whether the particle has passed
through this or that of the open slits. This might give rise to the
objection that our random variables do not represent any measurements
factually performed, whereas in the traditional contextuality analysis,
e.g., in the Kochen-Specker paradigm \cite{KochenSpecker1967}, the
random variables always represent results of measurements. This objection
would have a merit if the contextuality analysis of the double-slit
experiment represented by our random variables $R_{q}^{c}$ led to
different conclusions depending on what unobservable scenarios are
considered, and if there were no ways of determining, at least in
principle, comparative plausibility of different scenarios. We will
see, however, that the double-slit system in our analysis turns out
to be always noncontextual, making the question of physical plausibility
moot. The objection that the events like ``the particle passed (did
not pass) through this slit'' simply do not exist unless measured
would be a philosophical disagreement, discussing which here would
be out of place. It can be mentioned, however, that the very meaningfulness
of closing and opening slits in this experiment is contingent upon
one's believing that something related to the emitted particle somehow
passes or fails to pass through these slits. It is not unreasonable
therefore to ascribe physical meaning to our random variables, even
if not measured. 

In Section \ref{subsec: A-glimpse} we show that with the triple-slit
experiment the situation is different: there, for any nonzero probabilities
of the particle eventually hitting the detector in different contexts,
one can construct both a contextual scenario and a noncontextual scenario
of a particle passing through this or that of the open slits. Because
of this, in the absence of physical considerations constraining these
scenarios, we consider this result as only ``a glimpse'' into the
triple-slit system, subject to further speculations if not testing. 

Contextuality or noncontextuality is a property of a system of random
variables representing an empirical situation rather than of the empirical
situation itself. Our contextuality analysis pertains to our specific
choice of the random variables $R_{q}^{c}$, and it seems it has not
been explored previously. There are, however, other possible representations
of the double and triple-slit experiments by systems of random variables,
and one of them, unrelated to ours, has been considered and will be
mentioned in the concluding section. 

\section{Preliminaries: Contextuality-by-Default approach}

The departure point of CbD analysis is representing an empirical situation
as a \emph{content-context system} of random variables. This is a
set of random variables $R_{q}^{c}$ each of which is labeled by its
\emph{content} $q$, which means, roughly, that which the random variable
``measures'' or ``responds to,'' and its \emph{context} $c$,
the circumstances under which this measurement is made, including
but not limited to other contents measured together with a given one.
By construction, random variables sharing a context, $R_{q_{1}}^{c},\ldots,R_{q_{k}}^{c}$,
always have a uniquely defined \emph{joint distribution} (they are
measured ``together''), while any two random variables in different
(hence mutually exclusive) contexts, $R_{q}^{c}$ and $R_{q'}^{c'}$,
are \emph{stochastically unrelated}. In particular, in CbD, random
variables in different contexts can never be the same.\footnote{This allows one to avoid the logical problem one encounters in traditional
treatments of contextuality, where the sets of random variables in
different contexts have nonempty intersections. The problem arises
from two facts: (1) any contextuality analysis aims at establishing
the existence or non-existence of certain joint distributions, understood
in the classical (Kolmogorovian) sense; (2) in classical probability
theory the relation of being jointly distributed is transitive. The
conjunction of these two facts makes internally contradictory any
claim that, say, a joint distribution of $A,B,C$ does not exist while
the pairs $A,B$ and $B,C$ possess joint distributions. For detailed
discussion, see Ref. \cite{DzhKuj2017Fortsch}. } 

A \emph{(probabilisitic) coupling} of the system of random variables
is a set of jointly distributed random variables $S_{q}^{c}$, in
a one-to-one correspondence with $R_{q}^{c}$, such that the joint
distribution of any subset of context-sharing $S_{q_{1}}^{c},\ldots,S_{q_{k}}^{c}$
is the same as that of $R_{q_{1}}^{c},\ldots,R_{q_{k}}^{c}$. In the
traditional analysis of contextuality, the system of random variables
$R_{q}^{c}$ is assumed to be \emph{consistently connected}, which
means that any two random variables sharing a content, $R_{q}^{c}$
and $R_{q}^{c'}$, are identically distributed (while being distinct
and stochastically unrelated). The condition of consistent connectedness
is known in physics under the names of ``no-disturbance,'' ``no-signaling,''
etc. \cite{Kurzynski2014,Cereceda2000} The traditional definition
of a \emph{noncontextual system }of random variables $R_{q}^{c}$,
formulated in the language of CbD, is that this is a system that has
a coupling in which $S_{q}^{c}=S_{q}^{c'}$ holds with probability
1 for any two content-sharing random variables $R_{q}^{c}$ and $R_{q}^{c'}$.
Such a coupling need not exist, and if it does not, the system is
\emph{contextual}.

The problem with this definition (and the main motivation behind CbD,
beside the need of reconciling contextuality with rigorous probability
theory) is that any \emph{inconsistently connected} system of random
variables (one in which the distributions of $R_{q}^{c}$ and $R_{q}^{c'}$
may differ) is then ``automatically'' rendered contextual or else
placed outside the sphere of applicability of the notion of (non)contextuality.
Both these ways of treating inconsistent connectedness, while logically
valid, trivialize and severely restrict contextuality analysis. Consistent
connectedness is often violated in quantum physics, and it is virtually
nonexistent in non-physical applications. Thus, in Ref. \cite{KujDzhLar2015}
we re-analyze an experiment \cite{Lapkiewicz2011} exhibiting inconsistent
connectedness in the Klyachko-Can-Binicio\u{g}lu-Shumvosky paradigm
\cite{Klyachko2008}. In the Bohm-Aharonov version of the Einstein-Podolsky-Rosen
(EPR) entanglement paradigm \cite{Bohm1957}, famously investigated
by Bell and others \cite{Bell1964,Bell1966,CHSH1969,Fine1982}, consistent
connectedness is theoretically ensured by space-like separation of
the entangled particles. However, in real experiments inconsistency
is often present due to systematic design biases \cite{Adenier-Khrennikov 2007}.
The two particles may also be time-like separated in some experiments,
in which case inconsistent connectedness may be due to factual signaling
between the particles \cite{Bacon-Toner 2003}. In the Leggett-Garg
paradigm \cite{LeggettGarg1985}, later measurements may very well
be directly affected by the previous settings (\textquotedblleft signaling
in time,\textquotedblright{} \cite{Kofler2013,Budroni2013,Budroni2014,BudroniBook2016}),
and Bacciagaluppi systematically investigated the ensuing inconsistent
connectedness using the CbD approach \cite{Bacciagaluppi,Guido2016}.
In behavioral applications, there were several attempts to demonstrate
contextuality analogous to the EPR-Bell or Leggett-Garg systems, all
these attempts being frustrated by the ubiquity of inconsistent connectedness
in behavioral systems (for detailed analysis, see Refs. \cite{DZK2016,DKCZJ2016,CervDzh2018}). 

Intuitively, inconsistent connectedness is a manifestation of direct
causal action of experimental set-up upon the variables measured in
it (hence the terminology of ``disturbance,'' ``invasiveness,''
etc.). Contextuality, by contrast, is of a correlational, non-causal
nature: even if $R_{q}^{c}$ and $R_{q}^{c'}$ are identically distributed,
their correlations with other random variables in the respective contexts
make it impossible to map them into two always-equal $S_{q}^{c}$
and $S_{q}^{c'}$ within a coupling. In other words, the difference
in the \emph{identities} of the two random variables cannot be explained
by the difference of their distributions (in this case, no difference).\footnote{A random variable is \emph{identified} as a measurable function from
a probability space into a measurable space. \emph{Distribution} (the
measure induced by this mapping in the codomain space) is only one
aspect of the random variable's identity.} It seems reasonable therefore to extend the definition of contextuality
to allow (non)contextuality and (in)consistent connectedness to coexist
in all four possible combinations. In CbD, this is achieved by considering
the maximal possible probability with which jointly distributed $S_{q}^{c}$
and $S_{q}^{c'}$ (having the same individual distributions as $R_{q}^{c}$
and $R_{q}^{c'}$, respectively) can be equal to each other. This
probability equals 1 if the two distributions are the same, and if
they are not, it is viewed as a measure of the difference between
them. The question of contextuality then is translated into whether
this difference in distributions is sufficient to account for the
difference between the random variables' identities:
\begin{description}
\item [{Q2:}] given (generally different) distributions of $R_{q}^{c}$
and $R_{q}^{c'}$, do their correlations with other random variables
in their respective contexts make it possible to map them into jointly
distributed $S_{q}^{c}$ and $S_{q}^{c'}$ (within a coupling of the
system containing $R_{q}^{c}$ and $R_{q}^{c'}$) that are equal to
each other with the maximal possible probability?
\end{description}
The main idea underlying CbD is that if this question is answered
in the affirmative for every pair of content-sharing random variables,
the system is noncontextual. Otherwise it is contextual. An important
initial step in the analysis is that each random variable in the system
is to be dichotomized, replaced by a set of binary variables, for
reasons discussed in Refs. \cite{DzhCerKuj2017,DzhKuj2017Fortsch,Dzh2017Nothing}.\footnote{In a nutshell, a system of random variables amenable to contextuality
analysis should satisfy certain desiderata, such as uniqueness of
the coupling for any set of content-sharing random variables, and
the preservation of noncontextuality under deletions and coarse-graining
of the random variables.} We skip this discussion, as the system to be dealt with in this paper
consists of random variables that are already binary. As this system
turns out to be noncontextual, we also skip the otherwise important
issue of measuring the \emph{degree of contextuality} in systems found
to be contextual \cite{DzhKujLar2015,DzhKuj2016MathPsych}. 

\section{The content-context representation of the double-slit experiment}

\begin{figure}
\fbox{\begin{minipage}[t]{0.5\columnwidth}%
\includegraphics[scale=0.25]{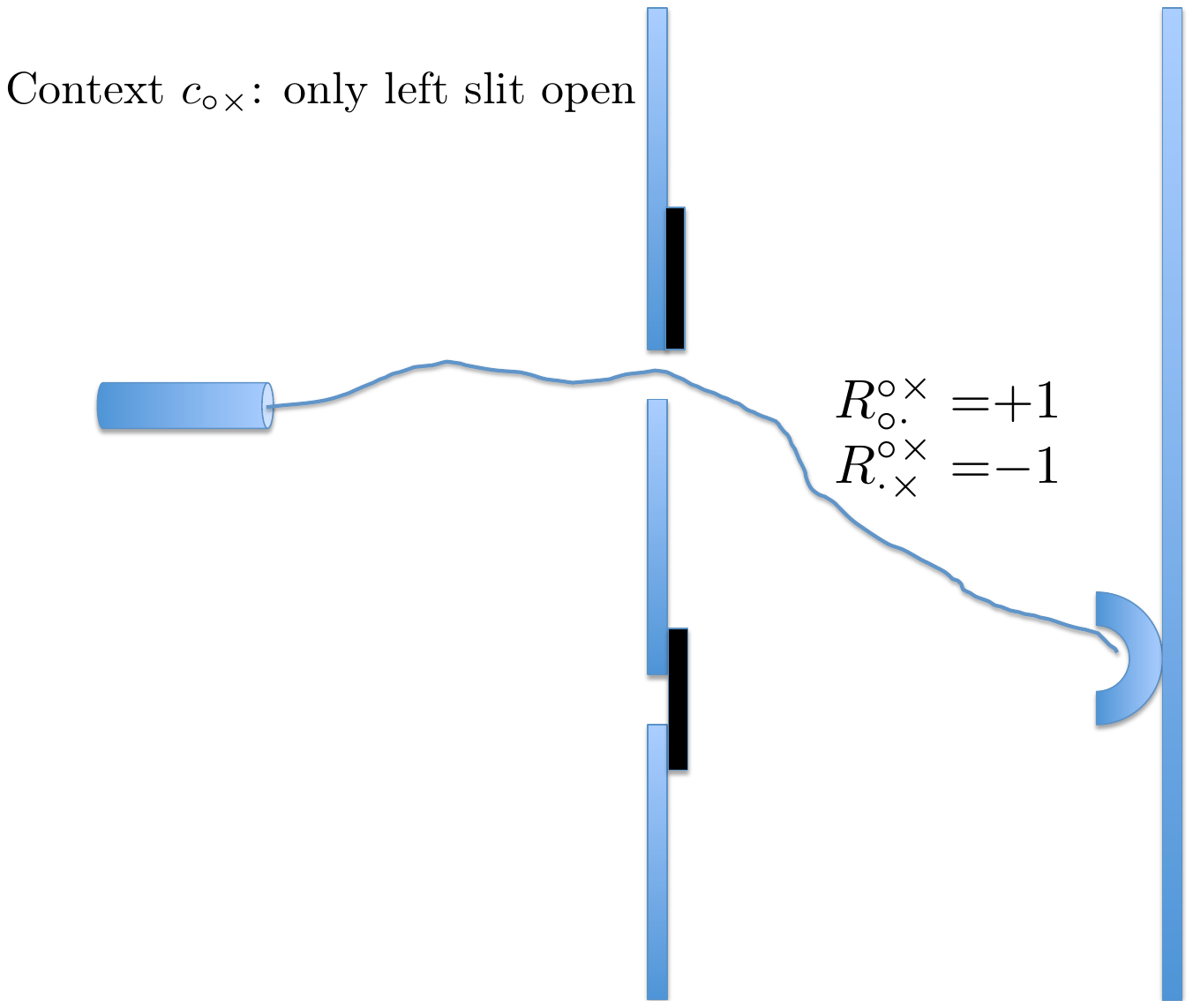}%
\end{minipage}}%
\fbox{\begin{minipage}[t]{0.5\columnwidth}%
\includegraphics[scale=0.25]{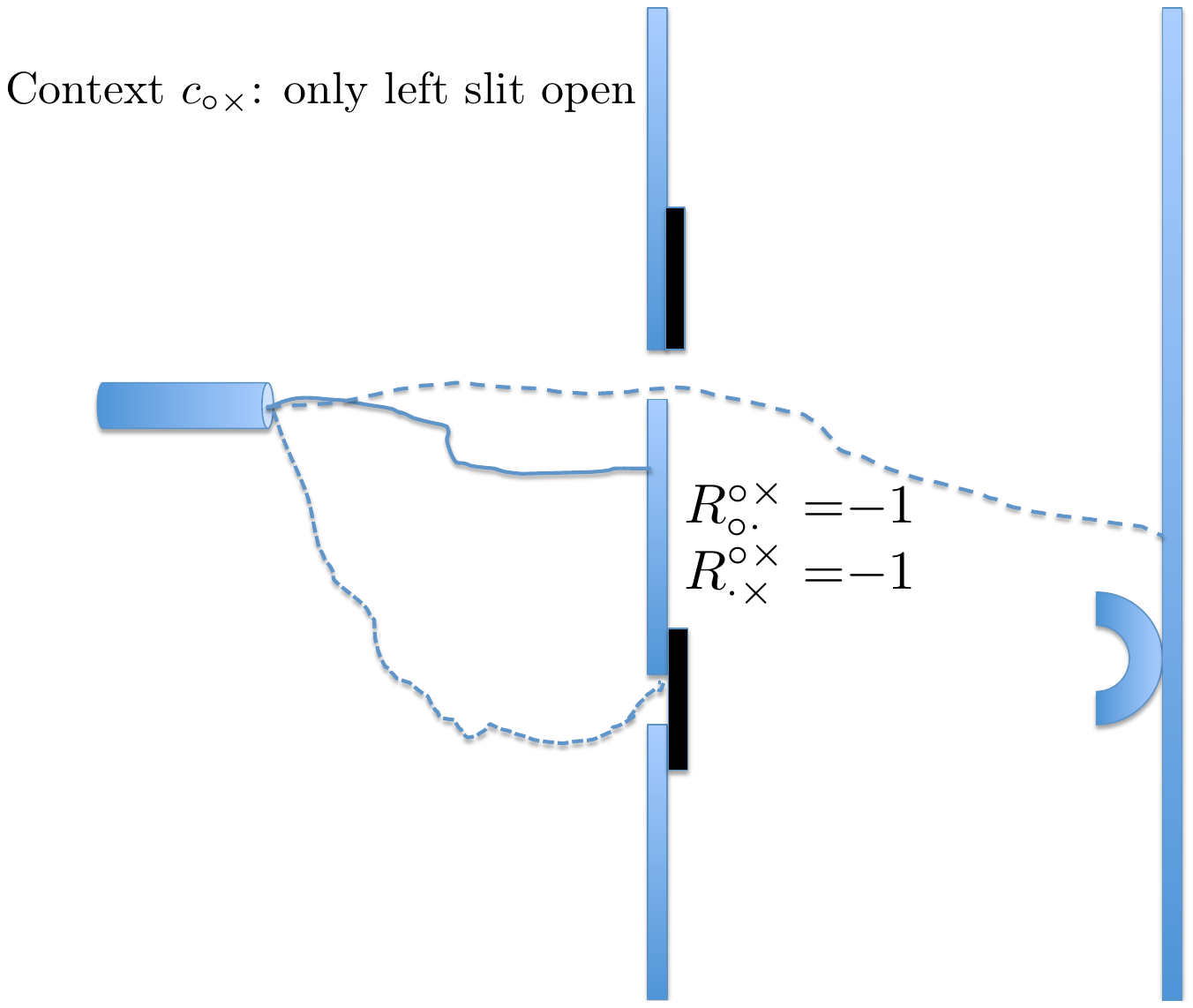}%
\end{minipage}}

\caption{\label{fig: cox}The only physical constraint adopted in our analysis
is that a particle can only hit the detector if it has passed through
an open slit. The random variables in context $c_{\circ\times}$ are
$R_{\circ\cdot}^{\circ\times}$ (has the particle passed through the
open left slit and hit the detector?) and $R_{\cdot\times}^{\circ\times}$
(has the particle passed through the closed right slit and hit the
detector?). The physical constraint we impose implies that $R_{\cdot\times}^{\circ\times}=-1$
with probability 1, which in turn implies that in context $c_{\circ\times}$
the only outcomes that can have nonzero probabilities are as shown
in the two panels. The solid irregular curve shows a possible ``trajectory''
of a single particle, dashed lines show other possible ``trajectories.'' }
\end{figure}

\begin{figure}
\begin{centering}
\fbox{\begin{minipage}[t]{0.5\columnwidth}%
\includegraphics[scale=0.25]{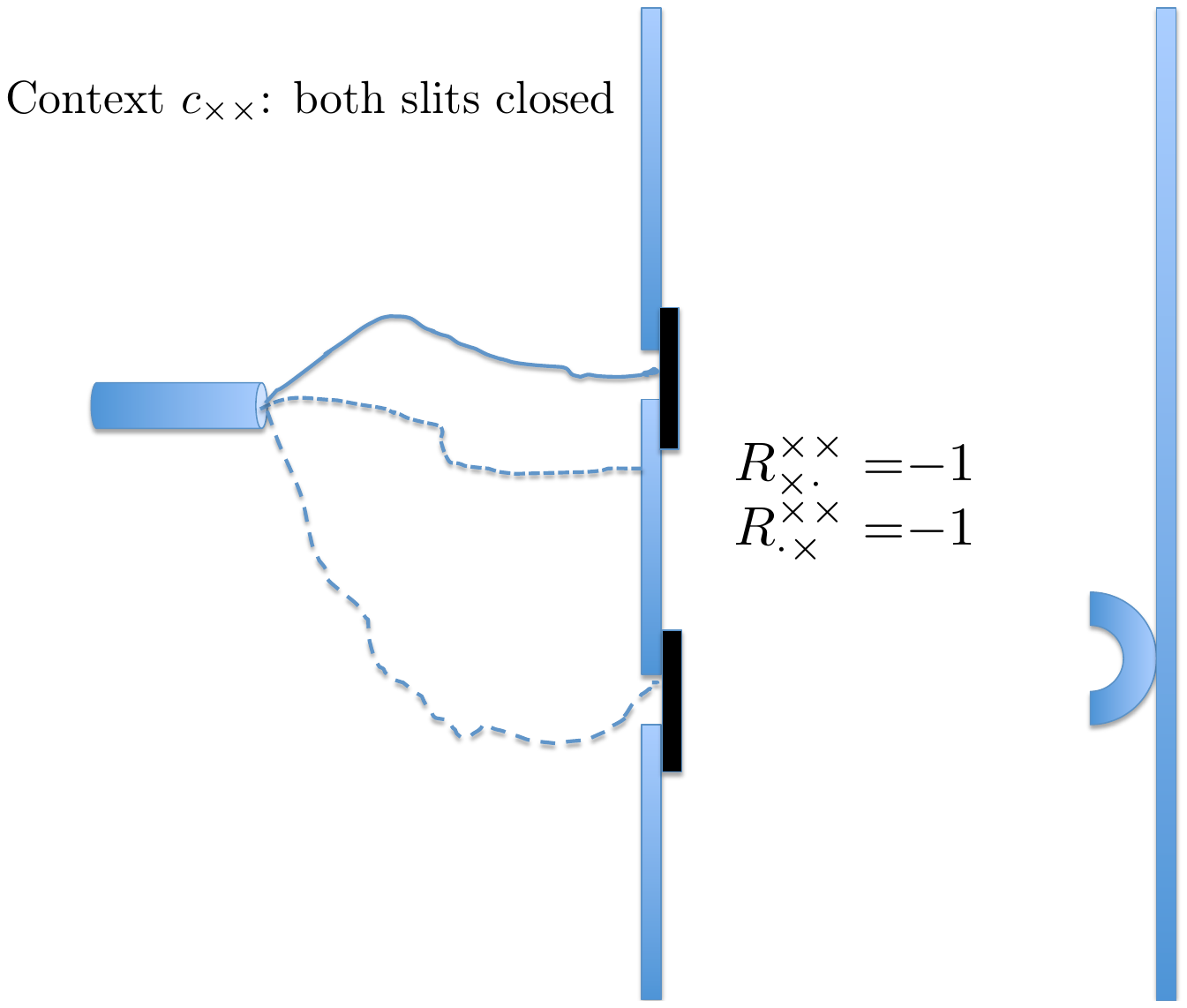}%
\end{minipage}}
\par\end{centering}
\caption{\label{fig: cxx}The two random variables in context $c_{\times\times}$
can only attain the values $-1$. The situation shown therefore occurs
with probability 1.}
\end{figure}

\begin{figure}
\fbox{\begin{minipage}[t]{0.5\columnwidth}%
\includegraphics[scale=0.25]{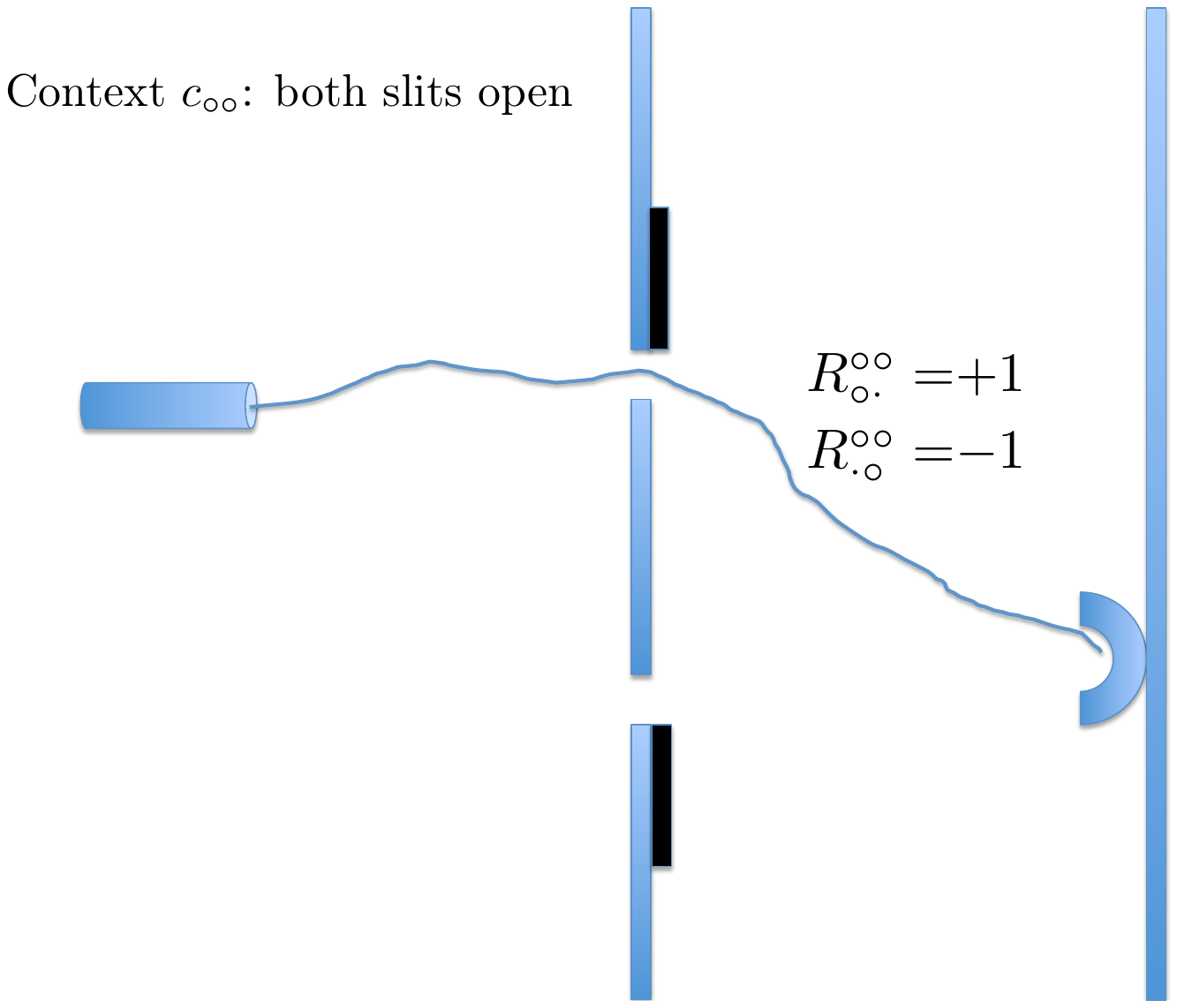}%
\end{minipage}}%
\fbox{\begin{minipage}[t]{0.5\columnwidth}%
\includegraphics[scale=0.25]{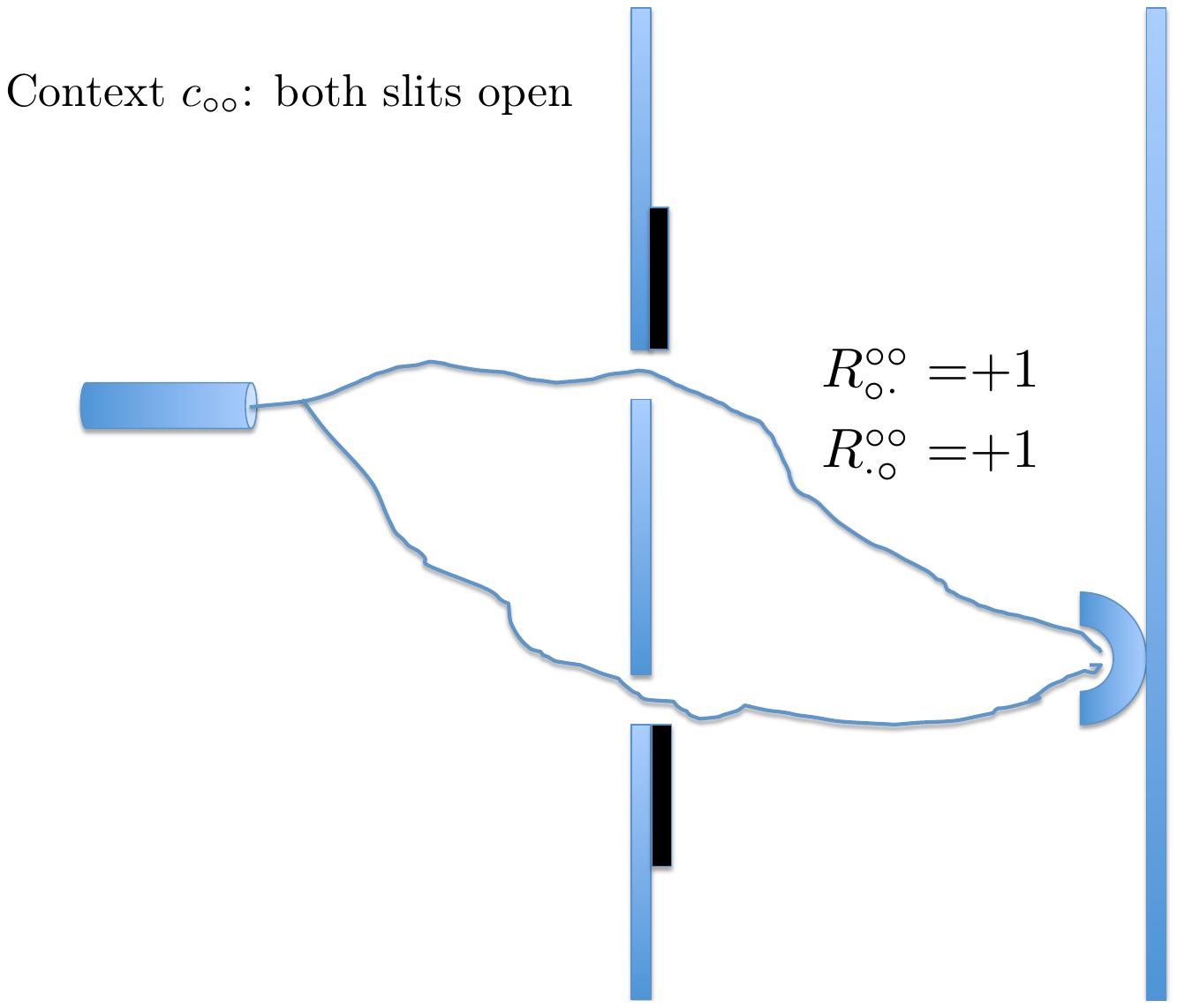}%
\end{minipage}}

\caption{\label{fig: coo}In context $c_{\circ\circ}$ a particle can hit the
detector having passed through one of the open slits (left panel),
but we also allow for the possibility that it passes through both
slits (right panel), or passes through them several times (not shown).
The reason for this is that the outcome of the contextuality analysis
does not depend on the probability of the situation in the right panel.}
\end{figure}

The \emph{content-context system} of the random variables we have
chosen to represent the double-slit experiment is
\begin{equation}
\begin{array}{|c|c|c|c||c|}
\hline R_{\circ\cdot}^{\circ\circ} & R_{\cdot\circ}^{\circ\circ} &  &  & c_{\circ\circ}\\
\hline  & R_{\cdot\circ}^{\times\circ} & R_{\times\cdot}^{\times\circ} &  & c_{\times\circ}\\
\hline  &  & R_{\times\cdot}^{\times\times} & R_{\cdot\times}^{\times\times} & c_{\times\times}\\
\hline R_{\circ\cdot}^{\circ\times} &  &  & R_{\cdot\times}^{\circ\times} & c_{\circ\times}\\
\hline\hline q_{\circ\cdot} & q_{\cdot\circ} & q_{\times\cdot} & q_{\cdot\times} & \textnormal{system SS}
\\\hline \end{array}\label{eq: matrix c-c}
\end{equation}
The superscripts of the random variables show their contexts, the
left (right) symbol indicating whether the left (right) slit is open
($\circ$) or closed ($\times$). The subscript of each random variable
shows which of these two slits is being ``measured'' by this random
variable, i.e., about which slit we ask the question Q1: e.g., $\circ\cdot$
in $R_{\circ\cdot}^{\circ\times}$ shows that the question is being
asked about the left open slit (when the right one is closed). The
random variables can be arranged in the following cyclic structure:
\begin{equation}
\vcenter{\xymatrix@C=1cm{ & R_{\circ\cdot}^{\circ\times}\ar@{-}[r]_{c_{\circ\times}} & R_{\cdot\times}^{\circ\times}\ar@{.}[dr]^{q_{\cdot\times}}\\
R_{\circ\cdot}^{\circ\circ}\ar@{.}[ur]^{q_{\circ\cdot}} &  &  & R_{\cdot\times}^{\times\times}\ar@{-}[d]_{c_{\times\times}}\\
R_{\cdot\circ}^{\circ\circ}\ar@{-}[u]_{c_{\circ\circ}} &  &  & R_{\times\cdot}^{\times\times}\ar@{.}[dl]^{q_{\times\cdot}}\\
 & R_{\cdot\circ}^{\times\circ}\ar@{.}[lu]^{q_{\cdot\circ}} & R_{\times\cdot}^{\times\circ}\ar@{-}[l]_{c_{\times\circ}}
}
}\label{eq: diagram}
\end{equation}
According to CbD, the random variables sharing a context, i.e., those
in the same row of matrix (\ref{eq: matrix c-c}) and connected by
solid lines in the diagram (\ref{eq: diagram}), are jointly distributed.
Let us present these distributions with references to the corresponding
graphical illustrations:
\begin{equation}
\textnormal{(see Fig. \ref{fig: cox})}\quad\begin{array}{c|c|c|c}
\textnormal{context }c_{\circ\times} & R_{\cdot\times}^{\circ\times}=+1 & R_{\cdot\times}^{\circ\times}=-1\\
\hline R_{\circ\cdot}^{\circ\times}=+1 & 0 & p & p\\
\hline R_{\circ\cdot}^{\circ\times}=-1 & 0 & \mathbf{1-p} & 1-p\\
\hline  & 0 & 1
\end{array}\label{eq: cox}
\end{equation}
\begin{equation}
\textnormal{(see Fig. \ref{fig: cxx})}\quad\begin{array}{c|c|c|c}
\textnormal{context }c_{\times\times} & R_{\cdot\times}^{\times\times}=+1 & R_{\cdot\times}^{\times\times}=-1\\
\hline R_{\times\cdot}^{\times\times}=+1 & 0 & 0 & 0\\
\hline R_{\times\cdot}^{\times\times}=-1 & 0 & \mathbf{1} & 1\\
\hline  & 0 & 1
\end{array}
\end{equation}

\begin{equation}
\begin{array}{c}
\textnormal{(same as Fig. \ref{fig: cox}}\\
\textnormal{with left and}\\
\textnormal{ right reversed)}
\end{array}\textnormal{}\quad\begin{array}{c|c|c|c}
\textnormal{context }c_{\times\circ} & R_{\cdot\circ}^{\times\circ}=+1 & R_{\cdot\circ}^{\times\circ}=-1\\
\hline R_{\times\cdot}^{\times\circ}=+1 & 0 & 0 & 0\\
\hline R_{\times\cdot}^{\times\circ}=-1 & q & \mathbf{1-q} & 1\\
\hline  & 0 & 1
\end{array}
\end{equation}
\begin{equation}
\textnormal{(see Fig. \ref{fig: coo})}\quad\begin{array}{c|c|c|c}
\textnormal{context }c_{\circ\circ} & R_{\cdot\circ}^{\circ\circ}=+1 & R_{\cdot\circ}^{\circ\circ}=-1\\
\hline R_{\circ\cdot}^{\circ\circ}=+1 & \ensuremath{r'} & \ensuremath{p'} & r'+p'\\
\hline R_{\circ\cdot}^{\circ\circ}=-1 & q' & \ensuremath{\mathbf{1-p'-q'-r'}} & 1-r'-p'\\
\hline  & r'+q' & 1-r'-q'
\end{array}\label{eq: coo}
\end{equation}
In each of these distributions, the only observable probability (i.e.,
one that can be estimated from empirical data) is the nondetection
probability shown in boldface. The rest of the probabilities represent
the scenarios of the particle passing through the open slits in all
imaginable ways.

\section{Contextuality analysis of the double-slit experiment}

According to CbD, the system shown in (\ref{eq: matrix c-c}) and
(\ref{eq: diagram}) is noncontextual if and only if one can find
eight jointly distributed random variables
\begin{equation}
\begin{array}{|c|c|c|c||c|}
\hline S_{\circ\cdot}^{\circ\circ} & S_{\cdot\circ}^{\circ\circ} &  &  & c_{\circ\circ}\\
\hline  & S_{\cdot\circ}^{\times\circ} & S_{\times\cdot}^{\times\circ} &  & c_{\times\circ}\\
\hline  &  & S_{\times\cdot}^{\times\times} & S_{\cdot\times}^{\times\times} & c_{\times\times}\\
\hline S_{\circ\cdot}^{\circ\times} &  &  & S_{\cdot\times}^{\circ\times} & c_{\circ\times}\\
\hline\hline q_{\circ\cdot} & q_{\cdot\circ} & q_{\times\cdot} & q_{\cdot\times} & \textnormal{coupling of SS}
\\\hline \end{array}\label{eq: coupling S}
\end{equation}
in one-to-one correspondence with the elements of (\ref{eq: matrix c-c}),
with the following properties:
\begin{enumerate}
\item The variables $S_{q}^{c}$ form a coupling of the system (\ref{eq: matrix c-c}).
This means that in each row of (\ref{eq: coupling S}) the random
variables have the same joint distribution as in the corresponding
row of (\ref{eq: matrix c-c}). 
\item This coupling is \emph{multimaximally connected}. This means that
in each column of (\ref{eq: coupling S}) the two random variables
have the joint distribution in which they equal to each other with
the maximal possible probability. This maximal probability is constrained
by the individual distributions of the two random variables, coinciding
with those of the corresponding variables in (\ref{eq: matrix c-c}).
(If the system contained more than two variables in a column, this
maximality requirement would be applied to every pair of them. With
all variables binary, this requirement can always be satisfied and
in precisely one possible way \cite{DzhKuj2017Fortsch,DzhCerKuj2017}.)
\end{enumerate}
The first requirement is simple: all probabilities shown in (\ref{eq: cox})-(\ref{eq: coo})
remain unchanged if one replaces each $R_{q}^{c}$ in them with the
corresponding $S_{q}^{c}$. To understand the second requirement,
consider, e.g., the first column in (\ref{eq: coupling S}). The probability
of $S_{\circ\cdot}^{\circ\circ}=S_{\circ\cdot}^{\circ\times}$ is
the sum of $\textnormal{Prob}\left[S_{\circ\cdot}^{\circ\circ}=S_{\circ\cdot}^{\circ\times}=1\right]$
and $\textnormal{Prob}\left[S_{\circ\cdot}^{\circ\circ}=S_{\circ\cdot}^{\circ\times}=-1\right]$,
and their maximal possible values are
\begin{equation}
\begin{array}{c}
\textnormal{Prob}\left[S_{\circ\cdot}^{\circ\circ}=1,S_{\circ\cdot}^{\circ\times}=1\right]=\min\left(\textnormal{Prob}\left[S_{\circ\cdot}^{\circ\circ}=1\right],\textnormal{Prob}\left[S_{\circ\cdot}^{\circ\times}=1\right]\right),\\
\\
\textnormal{Prob}\left[S_{\circ\cdot}^{\circ\circ}=-1,S_{\circ\cdot}^{\circ\times}=-1\right]=\min\left(\textnormal{Prob}\left[S_{\circ\cdot}^{\circ\circ}=-1\right],\textnormal{Prob}\left[S_{\circ\cdot}^{\circ\times}=-1\right]\right).
\end{array}
\end{equation}
This determines the joint probability of $\left(S_{\circ\cdot}^{\circ\circ},S_{\circ\cdot}^{\circ\times}\right)$
uniquely. Using the probabilities shown in (\ref{eq: cox}) and (\ref{eq: coo}),
\begin{equation}
\begin{array}{c|c|c|c}
\textnormal{content }q_{\circ\cdot} & S_{\circ\cdot}^{\circ\times}=+1 & S_{\circ\cdot}^{\circ\times}=-1\\
\hline S_{\circ\cdot}^{\circ\circ}=+1 & \min\left(p,r'+p'\right) & r'+p'-\min\left(p,r'+p'\right) & r'+p'\\
\hline S_{\circ\cdot}^{\circ\circ}=-1 & p-\min\left(p,r'+p'\right) & \min\left(1-p,1-r'-p'\right) & 1-r'-p'\\
\hline  & p & 1-p
\end{array}
\end{equation}
The joint distributions for the remaining three contents (columns)
of (\ref{eq: coupling S}) are computed similarly. 

We see therefore that in the hypothetical coupling (\ref{eq: coupling S})
the distributions in each row and in each column are uniquely specified.
The question of whether the system (\ref{eq: matrix c-c}) is (non)contextual
becomes the question of whether these row-wise and column-wise distributions
in (\ref{eq: coupling S}) are mutually compatible, i.e., whether
there is a joint distribution of all eight random variables in (\ref{eq: coupling S})
with these row-wise and column-wise distributions as its marginals.
Our system of random variables (\ref{eq: matrix c-c})-(\ref{eq: diagram})
is a cyclic system of rank 4 \cite{DzhKujLar2015}, also used to describe
the EPR-Bell experiment with spin-$1/2$ particles \cite{Bell1966,CHSH1969,Fine1982}.
One can therefore answer the question about compatibility by using
the criterion of (non)contextuality of a cyclic system derived in
Ref. \cite{KujDzh2016}. 

In general, a cyclic system of rank $n\geq2$ consists of $2n$ binary
random variables arranged so that each context $c_{i}$ ($i=1,\ldots,n$)
is defined by two contents $q_{i},q_{i\oplus1}$ measured together,
and each content $q_{i\oplus1}$ enters in two contexts $c_{i},c_{i\oplus1}$
(where $i\oplus1$ is simply $i+1$ except for $n\oplus1=1$). This
system is noncontextual (i.e., it has a multimaximally connected coupling)
if and only if
\begin{equation}
\max_{\left\{ \lambda_{1},\ldots,\lambda_{n}\right\} \in\Lambda_{n}}\sum_{i=1}^{n}\lambda_{i}\left\langle R_{i}^{i}R_{i\oplus1}^{i}\right\rangle \leq n-2+\sum_{i=1}^{n}\left|\left\langle R_{i\oplus1}^{i}\right\rangle -\left\langle R_{i\oplus1}^{i\oplus1}\right\rangle \right|,\label{eq: criterion for cyclic n}
\end{equation}
where $\Lambda_{n}$ denote the set of $n$-tuples $\left\{ \lambda_{1},\ldots,\lambda_{n}\right\} $
such that $\lambda_{i}\in\left\{ -1,+1\right\} $ and $\prod_{i=1}^{n}\lambda_{i}=-1$
(i.e., the number of the minus signs in the left-hand side sum is
odd). If the system is consistently connected, the sum of $\left|\left\langle R_{i\oplus1}^{i}\right\rangle -\left\langle R_{i\oplus1}^{i\oplus1}\right\rangle \right|$
in the right-hand side disappears, and the criterion coincides with
the one derived (in a very different way) in Ref. \cite{Araujoetal2013}. 

By simple if tedious algebra the expected values entering (\ref{eq: criterion for cyclic n})
can be computed for $n=4$ using (\ref{eq: cox})-(\ref{eq: coo}),
and the result is that (\ref{eq: criterion for cyclic n}) is satisfied
irrespective of the probability values in (\ref{eq: cox})-(\ref{eq: coo}).
The double-slit experiment represented by our random variables (\ref{eq: matrix c-c})-(\ref{eq: diagram})
is always noncontextual. 

There is, however, a much simpler way of establishing this noncontextuality.
In the matrix (\ref{eq: matrix c-c}) the random variables with contents
$q_{\times\cdot}$ and $q_{\cdot\times}$ are deterministic (equal
to $-1$ with probability 1). As shown in Ref. \cite{Dzh2017Nothing},
adding or deleting a deterministic quantity to/from a system of random
variables does not change its contextuality or noncontextuality.\footnote{\label{fn: degree1}In fact, the statement is stronger: the system's
\emph{degree of contextuality} does not change. We do not discuss
this notion here.} The system therefore is equivalent (with respect to its contextuality)
to

\begin{equation}
\begin{array}{|c|c||c|}
\hline R_{\circ\cdot}^{\circ\circ} & R_{\cdot\circ}^{\circ\circ} & c_{\circ\circ}\\
\hline  & R_{\cdot\circ}^{\times\circ} & c_{\times\circ}\\
\hline  &  & c_{\times\times}\\
\hline R_{\circ\cdot}^{\circ\times} &  & \\
\hline\hline q_{\circ\cdot} & q_{\cdot\circ} & \textnormal{system SS}'
\\\hline \end{array}\;.
\end{equation}
It is also clear that deleting a context containing just one or no
random variables does not change the system's contextuality or noncontextuality.\footnote{The statement in footnote \ref{fn: degree1} applies here too.}
The system therefore can be replaced with

\begin{equation}
\begin{array}{|c|c||c|}
\hline R_{\circ\cdot}^{\circ\circ} & R_{\cdot\circ}^{\circ\circ} & c_{\circ\circ}\\
\hline\hline q_{\circ\cdot} & q_{\cdot\circ} & \textnormal{system SS}''
\\\hline \end{array}\;.
\end{equation}
whose noncontextuality is trivially apparent.

\section{\label{subsec: A-glimpse}A glimpse into the triple-slit system}

The noncontextuality of the double-slit system does not depend on
whether it is physically realizable: it holds for any system (\ref{eq: matrix c-c}).
The situation with systems with three or more slits is different.
Consider the triple-slit system

\begin{equation}
\begin{array}{|c|c|c|c|c|c||c|}
\hline  &  &  & R_{\cdot\cdot\times}^{\times\times\times} & R_{\cdot\times\cdot}^{\times\times\times} & R_{\times\cdot\cdot}^{\times\times\times} & c_{\times\times\times}\\
\hline R_{\circ\cdot\cdot}^{\circ\times\times} &  &  & R_{\cdot\cdot\times}^{\circ\times\times} & R_{\cdot\times\cdot}^{\circ\times\times} &  & c_{\circ\times\times}\\
\hline  & R_{\cdot\circ\cdot}^{\times\circ\times} &  & R_{\cdot\cdot\times}^{\times\circ\times} &  & R_{\times\cdot\cdot}^{\times\circ\times} & c_{\times\circ\times}\\
\hline  &  & R_{\cdot\cdot\circ}^{\times\times\circ} &  & R_{\cdot\times\cdot}^{\times\times\circ} & R_{\times\cdot\cdot}^{\times\times\circ} & c_{\times\times\circ}\\
\hline R_{\circ\cdot\cdot}^{\circ\times\circ} &  & R_{\cdot\cdot\circ}^{\circ\times\circ} &  & R_{\cdot\times\cdot}^{\circ\times\circ} &  & c_{\circ\times\circ}\\
\hline R_{\circ\cdot\cdot}^{\circ\circ\times} & R_{\cdot\circ\cdot}^{\circ\circ\times} &  & R_{\cdot\cdot\times}^{\circ\circ\times} &  &  & c_{\circ\circ\times}\\
\hline  & R_{\cdot\circ\cdot}^{\times\circ\circ} & R_{\cdot\cdot\circ}^{\times\circ\circ} &  &  & R_{\times\cdot\cdot}^{\times\circ\circ} & c_{\times\circ\circ}\\
\hline R_{\circ\cdot\cdot}^{\circ\circ\circ} & R_{\cdot\circ\cdot}^{\circ\circ\circ} & R_{\cdot\cdot\circ}^{\circ\circ\circ} &  &  &  & c_{\circ\circ\circ}\\
\hline\hline q_{\circ\cdot\cdot} & q_{\cdot\circ\cdot} & q_{\cdot\cdot\circ} & q_{\cdot\cdot\times} & q_{\cdot\times\cdot} & q_{\times\cdot\cdot} & \textnormal{system SSS}
\\\hline \end{array}
\end{equation}
Using the same shortcut reasoning as with the system (\ref{eq: matrix c-c}),
i.e., deleting the columns with deterministic variables and the rows
with no more than one random variable, this triple-slit system is
equivalent (with respect to its contextuality) to

\begin{equation}
\begin{array}{|c|c|c||c|}
\hline R_{\circ\cdot\cdot}^{\circ\times\circ} &  & R_{\cdot\cdot\circ}^{\circ\times\circ} & c_{\circ\times\circ}\\
\hline R_{\circ\cdot\cdot}^{\circ\circ\times} & R_{\cdot\circ\cdot}^{\circ\circ\times} &  & c_{\circ\circ\times}\\
\hline  & R_{\cdot\circ\cdot}^{\times\circ\circ} & R_{\cdot\cdot\circ}^{\times\circ\circ} & c_{\times\circ\circ}\\
\hline R_{\circ\cdot\cdot}^{\circ\circ\circ} & R_{\cdot\circ\cdot}^{\circ\circ\circ} & R_{\cdot\cdot\circ}^{\circ\circ\circ} & c_{\circ\circ\circ}\\
\hline\hline q_{\circ\cdot\cdot} & q_{\cdot\circ\cdot} & q_{\cdot\cdot\circ} & \textnormal{system SSS}'
\\\hline \end{array}\label{eq: for 3 slits}
\end{equation}
Let the nondetection probabilities (the only observable ones) in these
four contexts be denoted 
\begin{equation}
\begin{array}{c||c|c|c|c|}
\textnormal{context} & c_{\circ\times\circ} & c_{\circ\circ\times} & c_{\times\circ\circ} & c_{\circ\circ\circ}\\
\hline \begin{array}{c}
\textnormal{probability that}\\
\textnormal{no detection occurs}
\end{array} & p & q & r & s
\end{array}\label{eq: detection}
\end{equation}
One can always find a noncontextual scenario for these probabilities,
e.g., the following one, in which all but one random variable in each
context are deterministic: in context $c_{\circ\circ\circ}$,

\begin{equation}
\textnormal{Prob}\left[\begin{array}{c}
R_{\circ\cdot\cdot}^{\circ\circ\circ}=i\\
R_{\cdot\circ\cdot}^{\circ\circ\circ}=j\\
R_{\cdot\cdot\circ}^{\circ\circ\circ}=k
\end{array}\right]=\left\{ \begin{array}{lclc}
s & \textnormal{if} & \left(i,j,k\right)=\left(-1,-1,-1\right)\\
1-s & \textnormal{if} & \left(i,j,k\right)=\left(+1,-1,-1\right)\\
0 & \textnormal{if} & \textnormal{otherwise}
\end{array}\right.
\end{equation}
and, in the three remaining contexts,
\begin{equation}
\begin{array}{c}
\begin{array}{c|c|c|c}
c_{\circ\times\circ} & R_{\cdot\cdot\circ}^{\circ\times\circ}=+1 & R_{\cdot\cdot\circ}^{\circ\times\circ}=-1\\
\hline R_{\circ\cdot\cdot}^{\circ\times\circ}=+1 & 0 & 1-p & 1-p\\
\hline R_{\circ\cdot\cdot}^{\circ\times\circ}=-1 & 0 & p & p\\
\hline  & 0 & 1
\end{array}\\
\\
\begin{array}{c|c|c|c}
c_{\circ\circ\times} & R_{\cdot\circ\cdot}^{\circ\circ\times}=+1 & R_{\cdot\circ\cdot}^{\circ\circ\times}=-1\\
\hline R_{\circ\cdot\cdot}^{\circ\circ\times}=+1 & 0 & 0 & 0\\
\hline R_{\circ\cdot\cdot}^{\circ\circ\times}=-1 & 1-q & q & 1\\
\hline  & 1-q & q
\end{array}\\
\\
\begin{array}{c|c|c|c}
c_{\times\circ\circ} & R_{\cdot\cdot\circ}^{\times\circ\circ}=+1 & R_{\cdot\cdot\circ}^{\times\circ\circ}=-1\\
\hline R_{\cdot\circ\cdot}^{\times\circ\circ}=+1 & 0 & 1-r & 1-r\\
\hline R_{\cdot\circ\cdot}^{\times\circ\circ}=-1 & 0 & r & r\\
\hline  & 0 & 1
\end{array}
\end{array}
\end{equation}

The nondetection probabilities (\ref{eq: detection}) are also compatible
with contextual scenarios, with some exceptions, e.g., if any three
of them equal 1. Not to deal with special cases, we construct a contextual
scenario under the additional assumption that $s<1$ and $p<1$ (where
$p$ can be replaced with $q$ or $r$). Choose a probability $t>\max\left(s,p\right)$
and put
\begin{equation}
\textnormal{Prob}\left[\begin{array}{c}
R_{\circ\cdot\cdot}^{\circ\circ\circ}=i\\
R_{\cdot\circ\cdot}^{\circ\circ\circ}=j\\
R_{\cdot\cdot\circ}^{\circ\circ\circ}=k
\end{array}\right]=\left\{ \begin{array}{lclc}
s & \textnormal{if} & \left(i,j,k\right)=\left(-1,-1,-1\right)\\
t-s & \textnormal{if} & \left(i,j,k\right)=\left(-1,+1,-1\right)\\
1-t & \textnormal{if} & \left(i,j,k\right)=\left(+1,+1,+1\right)\\
0 & \textnormal{if} & \textnormal{otherwise}
\end{array}\right.
\end{equation}
Consider the subsystem
\begin{equation}
\begin{array}{|c|c||c|}
\hline R_{\circ\cdot\cdot}^{\circ\times\circ} & R_{\cdot\cdot\circ}^{\circ\times\circ} & c_{\circ\times\circ}\\
\hline R_{\circ\cdot\cdot}^{\circ\circ\circ} & R_{\cdot\cdot\circ}^{\circ\circ\circ} & c_{\circ\circ\circ}\\
\hline\hline q_{\circ\cdot\cdot} & q_{\cdot\cdot\circ} & \textnormal{system SSS}''
\\\hline \end{array}
\end{equation}
of the system (\ref{eq: for 3 slits}). Define the two row-wise distributions
as 
\begin{equation}
\begin{array}{c}
\begin{array}{c|c|c|c}
c_{\circ\times\circ} & R_{\cdot\cdot\circ}^{\circ\times\circ}=+1 & R_{\cdot\cdot\circ}^{\circ\times\circ}=-1\\
\hline R_{\circ\cdot\cdot}^{\circ\times\circ}=+1 & 1-2t+p & t-p & 1-t\\
\hline R_{\circ\cdot\cdot}^{\circ\times\circ}=-1 & t-p & p & t\\
\hline  & 1-t & t
\end{array}\\
\\
\begin{array}{c|c|c|c}
c_{\circ\circ\circ} & R_{\cdot\cdot\circ}^{\circ\circ\circ}=+1 & R_{\cdot\cdot\circ}^{\circ\circ\circ}=-1\\
\hline R_{\circ\cdot\cdot}^{\circ\circ\circ}=+1 & 1-t & 0 & 1-t\\
\hline R_{\circ\cdot\cdot}^{\circ\circ\circ}=-1 & 0 & t & t\\
\hline  & 1-t & t
\end{array}
\end{array}
\end{equation}
This describes a consistently connected cyclic system of rank 2. The
contextuality of this subsystem (hence also the contextuality of the
entire system) can be verified by applying to it the criterion (\ref{eq: criterion for cyclic n})
with $n=2$, or simply observing that this system can be noncontextual
only if $t=p$.

\section{Concluding Remarks}

We have established that the system of random variables describing
the double-slit experiment (in terms of which open slits the particle
passes through before hitting or missing the detector) is noncontextual
for all possible scenarios. For experiments involving more than two
slits, the systems describing them can be contextual. In fact, excluding
some special cases, every set of observable (in the statistical sense)
detection probabilities in this case allows for a contextual scenario
and a noncontextual scenario. The interpretation of the noncontextuality
of the double-slit system is that all context-dependence in this system
is due to direct influences exerted by the state of a slit (open or
closed) upon the probabilities with which a particle passes through
the other slit and hits the detector. These direct influences are
manifested in the differences in the distributions of random variables
sharing a content (tied to the same open slit). By contrast, one can
construct triple-slit systems (on paper, their physical realizability
is open to investigation), in which the difference in the identity
of random variables tied to a given slit under different (open-closed)
arrangements of other slits cannot be accounted by the difference
in the distributions of these random variables alone: we have a ``pure
contextuality'' here, on top of any possible direct influences. Physical
mechanisms of direct influences play no role in our analysis. With
the exception of the prohibition for a particle to pass through a
closed slit, the analysis involves no physical assumptions whatever. 

As mentioned in the introductory part of the paper, contextuality
analysis characterizes a set of random variables rather than an empirical
situation that, while it can be described by this set of random variables,
allows for other descriptions. Our analysis pertains to a particular
choice of random variables, tying each of them to a particular slit
(left or right) in a particular state (open or closed). Each context
in our analysis involves two random variables in no particular chronological
relation to each other. Kofler and Brukner \cite{Kofler2013} explored
another way of looking at the double-slit experiment (more precisely,
at its simplified version provided by a Mach-Zehnder interferometer).
The contents there correspond to three chronological stages, $t_{1}$
(the stage preceding the first beam split), $t_{2}$ (between the
first and the second splits), and $t_{3}$ (following the second split).
With each of these stages one associates a binary random variable
whose values corresponds to the choice of one two possible paths.
The measurements are assumed to be made in pairs, $\left(t_{1},t_{2}\right)$,
$\left(t_{1},t_{3}\right)$, and $\left(t_{2},t_{3}\right)$, forming
three contexts. In the CbD language, this creates six contextually
labeled random variables forming a cyclic system of rank 3, essentially
the same as one used to describe the Leggett-Garg experiment \cite{LeggettGarg1985}.
Kofler and Brukner discuss contextuality of this system for the case
when it is consistently connected. Mansfield, in an unpublished conference
presentation \cite{key-19}, also discussed a cyclic-3 representation
for the double-slit experiment but in a more general version, allowing
for ``signaling in time.'' We see no obvious relations between these
analyses and ours, and they are only mentioned here for completeness. 

Richard Feynman is often cited as asserting that the double-slit experiment
is incompatible with classical probability. He characterized the interference
pattern as ``the discovery that in nature the laws of combining probabilities
were not those of the classical probability theory of Laplace\textquotedblright{}
(\cite{Feynman1951} p. 533), and he said that it is ``a phenomenon
which is impossible, \emph{absolutely} impossible, to explain in any
classical way'' (\cite{Feynmanetal1975}, Section 37-1). Although
one can think of alternative interpretations for these quotes and
find other quotes seemingly saying something else, this interpretation
is widely accepted (see, e.g., Refs. \cite{Ballentine1986,Accardi1982,Khren2009,Constantini1993}).
Our analysis contradicts this interpretation, whether historically
correct or not, as CbD is squarely an application of classical (Kolmogorovian)
probability theory. Feynman's claim (or alleged claim) has been challenged
by others as well, and all of these challenges were using some form
of contextual labeling of the random variables involved. Thus, Ballantine
\cite{Ballentine1986} and Khrennikov \cite{Khren2015TwoSlit,AvisFischerHilbertKhrennikov2009,Khren2015CHSH}
treat the probabilities $p,q,\ldots$ in matrices like our (\ref{eq: cox})-(\ref{eq: coo})
as \emph{conditional probabilities}, using the contexts as conditioning
random events. Even closer to CbD, Khrennikov \cite{Khre2009Book}
treats $p,q,\ldots$ as ``contextual probabilities,'' with $c_{\circ\times}$,
$c_{\times\times}$, being essentially labels rather than conditioning
events. In all these and similar treatments the conditional labeling
is used to show that the classical probabilistic formulas claimed
to be violated by quantum-mechanical phenomena simply do not apply.
For instance, the additivity of probabilities of disjoint events,
thought by Feynman to be violated by the double-slit experiment, does
not apply because the union of the disjoint events and the events
themselves are conditioned (or ``contextualized'') by different
contexts. In CbD, this is definitely true, but this is only a departure
point for subsequent contextuality analysis \cite{DzhKuj2014PloS}. 

\section*{Acknowledgments}

This work has been partially supported by the AFOSR grant FA9550-14-1-0318.
The authors are grateful to Pawel Kurzynski, J. Acacio de Barros,
and Victor H. Cervantes for valuable discussions.

\end{document}